\documentclass[review,authoryear]{elsarticle}
\usepackage[utf8]{inputenc}
\usepackage[T1]{fontenc}
\usepackage{amsmath,amssymb,amsthm,bm,latexsym,amsfonts}
\usepackage{lscape}
\usepackage{longtable}
\usepackage{booktabs}
\usepackage[margin=0.85in]{geometry}
\usepackage{float}  
\usepackage{hyperref}
\hypersetup{colorlinks=true,citecolor=blue,pdfpagemode=FullScreen,linktocpage=true}
\usepackage{cleveref}
\usepackage{fancyhdr}
\fancypagestyle{elsarticle}{
    \fancyhf{}
    
    \fancyfoot[L]{\thepage}
}
\usepackage{tabularx}
\usepackage{threeparttable}
\usepackage{adjustbox}
\usepackage{subcaption}
\usepackage{pdflscape}
\usepackage{float}
\usepackage{caption}
\usepackage{multirow}
\usepackage{etoolbox}
\usepackage{graphicx,psfrag,epsf}
\usepackage{epigraph,xcolor}
\usepackage{tikz}

\usepackage{ragged2e}
\usepackage{enumerate}
\usepackage{url}
\usepackage{orcidlink} 
\usepackage[toc]{appendix}

\begin{document}

\makeatletter
 
\patchcmd{\MaketitleBox}{\normalsize\elsauthors}{\Large\elsauthors}{}{}
    \makeatother
\begin{frontmatter}

\title{Assessing the Role of Green Premium and Localized Price Spillovers in Budget Hotels: Evidence from Delhi National Capital Region (NCR)}

\author[aff1]{Shreyash Deshmukh}

\author[aff1]{Shovan Chowdhury\texorpdfstring{\corref{cor1}}{}}
\ead{shovanc@iimk.ac.in}

\cortext[cor1]{Corresponding author}

\affiliation[aff1]{
    organization={Decision Sciences and Operations Management Area,
    Indian Institute of Management Kozhikode},
    addressline={IIMK Campus},
    city={Kozhikode},
    postcode={673570},
    state={Kerala},
    country={India}
}
\begin{abstract}
  Sustainable urban tourism depends on how environmental quality, public transport accessibility, and
  accommodation clustering shape visitor mobility and destination development. This study examines these
  relationships in the budget hotel sector of Delhi National Capital Region, India, an emerging mega-destination
  marked by rapid urban expansion and uneven infrastructure provision. Using data from 857 budget hotels, we
  estimate a mixed geographically weighted spatial autoregressive model (MGWR--SAR) that captures both spatial
  heterogeneity and localized price spillovers. The model combines hotel attributes with satellite-
  derived indicators, including vegetation density, built-up area, night-time light intensity, and
  distances to metro stations, bus stops, airports, city centers, and tourist attractions. Results show that
  greenery has a place dependent value: it generates a green premium in dense urban cores, where environmental
  amenities are scarce, but becomes a discount in peripheral areas where vegetation often reflects weaker
  tourism infrastructure. Metro accessibility creates a mobility premium, highlighting the role of public
  transport in supporting low-cost tourist movement. Local spatial spillovers reveal agglomeration benefits in
  established clusters and price cannibalization in peripheral markets. The findings show how spatially
  differentiated planning can align green infrastructure, transit investment, and budget accommodation with more
  sustainable urban tourism development.
  \end{abstract}

\begin{keyword}
Green Premium, Spatial Pricing Dynamics, Spatial Heterogeneity, Spatial Dependence, MGWR--SAR, Budget Hotels
\end{keyword} 
\end{frontmatter}

\section{Introduction}

In the tourism and hospitality sector, hotel prices reflect both consumer preferences and the value of bundled service attributes. The hedonic price model (HPM), a widely used revealed-preference approach, explains hotel room prices through intrinsic features and locational characteristics \citep{rosen1974hedonic}. Because the implicit price of any single attribute is not directly observable, the HPM estimates the marginal effects of multiple bundled attributes simultaneously. From a destination management and revenue management perspective, this framework helps identify which attributes matter most for pricing, product differentiation, and destination positioning \citep{vives2018revenue}.

Hotel pricing is shaped by a complex interaction of internal managerial decisions (e.g., brand strategy, room size, amenities, and discounts) and external factors (e.g., location, tourist accessibility, and destination environmental characteristics). In competitive tourism markets, price serves as a signal of value and quality, and hotel operators rely on dynamic pricing systems to optimize yields. While most applications of the HPM include a common set of core attributes such as room characteristics and user ratings, the specific valuation of local environmental features (such as urban greenery) and transport infrastructure remains relatively underexplored, particularly in the budget lodging segment. Guests of budget or limited-service hotels tend to be price-sensitive and exhibit low switching costs \citep{tanford2012determinants}. For destination management organizations (DMOs) and operators, this makes it critical to identify how spatial and environmental attributes influence pricing and competitive dynamics.

Traditional hedonic models assume that the relationship between hotel attributes and price is spatially uniform. In practice, this assumption is often violated due to spatial heterogeneity, where the marginal value of an attribute varies by location. For example, urban green spaces (vegetation density) may command a ``green premium'' in highly congested urban centers due to their aesthetic and environmental benefits, whereas in peripheral areas, abundant greenery might correspond to agricultural land and carry a discount. Similarly, hotel prices exhibit spatial dependence, where pricing decisions of neighboring hotels are highly interdependent due to spatial competition and demand spillovers. In tourism destinations, this often manifests as localized spatial spillovers or ``coopetition''---where hotels benefit from agglomeration economies in high-demand zones (cooperating to attract tourists to the cluster) but engage in intense price competition (undercutting neighbors) in peripheral or emerging sub-markets.

Conventional spatial models, such as the Spatial Autoregressive (SAR) and Spatial Error Models (SEM), address spatial dependence but assume a spatially constant relationship. Conversely, Geographically Weighted Regression (GWR) models spatial heterogeneity but assumes spatial independence of the residuals. Since local spatial spillovers and environmental valuation are localized and co-occur, a model that simultaneously captures both dimensions is necessary. This study integrates the GWR and SAR frameworks to implement a mixed geographically weighted spatial autoregressive model (MGWR--SAR) that captures both spatial heterogeneity and spatial dependence simultaneously.

This paper makes several contributions to the tourism and hospitality literature. First, we focus on the Delhi National Capital Region (NCR) in India, offering empirical insights from an emerging-market mega-destination characterized by rapid, multi-nodal development. Second, we integrate satellite-derived environmental and urban indicators---specifically, vegetation density (NDVI) to capture the green premium, and night-time light (NTL) intensity and built-up area to proxy economic activity---within a localized econometric framework. Third, by utilizing the MGWR--SAR model, we uncover the dual spatial forces of destination agglomeration and competitive price-undercutting across sub-markets. Our findings challenge spatial uniformity in tourist utility, providing policy-relevant insights for destination planning, public transit infrastructure, and spatial pricing strategies.

\section{Literature Review}

\subsection{Hedonic Pricing in Tourism and Hospitality}

The theoretical foundation for analyzing the price of multi-attribute goods is the hedonic pricing model (HPM), as established in the seminal works of \citet{rosen1974hedonic} and \citet{lancaster1966new}. The model posits that a product's price is determined by the sum of the implicit values of its individual characteristics. In the tourism and hospitality sector, the HPM is particularly useful because a hotel room is a bundle of diverse features, including physical characteristics, services, accessibility, and reputation. It allows researchers and revenue managers to deconstruct a room rate into the values of its constituent parts, helping to design hotel attributes based on a clear understanding of customers' willingness to pay for each feature \citep{soler2019algarve}.

Studies examining accommodation pricing across both traditional hotels and peer-to-peer platforms generally classify explanatory variables into a set of broad and consistent categories: structural (intrinsic) attributes, reputational or host characteristics, locational (extrinsic) attributes, rental policies, and temporal or market-related factors \citep{chica2020effects, faye2021methodological, sainaghi2022effects, soler2019algarve, wang2022contribution}. Structural or intrinsic attributes refer to the physical characteristics and quality signals of the property, such as the capacity, hotel star classification, and the range of amenities provided, including Wi-Fi, parking, air conditioning, and swimming pools \citep{chica2020effects, faye2021methodological, soler2019algarve, boto2021spatial, wang2022contribution, jiang2020hotel, fleischer2012room}. Reputational attributes, such as online reputation scores, review ratings, and review volume, reduce information asymmetry and help command price premiums by signaling quality \citep{faye2021methodological, sainaghi2022effects, chica2020effects, abrate2016strategic, wang2019tourism}. Finally, booking policies (e.g., cancellation rules, minimum stay) and temporal variations (seasonality, weekend effects, and reservation horizons) further shape dynamic pricing strategies in hotel markets \citep{boto2021spatial, sainaghi2022effects, faye2021methodological, wang2019tourism, juaneda2011pricing, soler2019algarve}.

\subsection{Agglomeration Economies and Spatial Pricing Dynamics in Tourism Destinations}

A defining feature of the tourism industry is spatial agglomeration, where hotels and tourism services cluster geographically to leverage external economies, such as shared tourist traffic, labor pools, public infrastructure, and localized destination brand equity \citep{yang2016market, deller2010rural}. This spatial concentration can give rise to a form of ``coopetition''---where hotels benefit collectively from clustering to attract visitors to a destination zone (cooperation), while simultaneously competing to capture bookings from the shared tourist pool (competition).

In spatial econometrics, these market interactions manifest as spatial dependence, where a hotel's pricing decisions are directly influenced by the prices of nearby properties. \citet{balaguer2013relationship} demonstrate that spatial agglomeration directly shapes price premiums, though its strength varies between business and leisure tourism segments. In prime destination centers, positive spatial spillovers occur because agglomeration benefits dominate: high tourist density supports stable prices across the cluster \citep{lee2011room}. Conversely, in peripheral or emerging zones, price cannibalization or competitive undercutting may prevail. When neighboring hotels reduce rates to attract price-sensitive budget tourists, nearby hotels must follow suit, leading to negative spatial spillovers \citep{abrate2016strategic}. Despite its theoretical importance, the dual nature of these spatial spillovers (positive and negative feedback) is rarely modeled locally across sub-markets.

\subsection{Transit Accessibility and Spatial Tourism Mobilities}

Locational convenience is a primary determinant of hotel pricing, and is closely tied to tourist mobility. Accessibility is traditionally measured by proximity to key urban amenities, such as city centers, airports, tourist attractions, and transport hubs \citep{yang2016market, chen2010application, deller2010rural, latinopoulos2018using, soler2019algarve, chica2020effects, wang2022contribution, lee2024spatial, rigall2011effects, juaneda2011pricing, balaguer2013relationship, lee2011room}. 

For budget travelers, who are highly price-sensitive, transport costs and spatial friction are critical considerations. In major metropolitan areas, proximity to public transit networks, such as metro and bus systems, acts as a primary value driver by facilitating low-cost mobilities. Proximity to transit hubs allows guests to easily navigate the city and access remote tourist sights, effectively expanding their spatial choice set. Studies show that hotels near major transport corridors and railway stations command a premium \citep{soler2018hedonic, wang2022contribution}. However, the valuation of accessibility is not spatially uniform; the convenience premium of transport nodes may vary depending on the destination's polycentric structure and localized traffic congestion.

\subsection{Environmental Valuations and the Green Tourism Premium}

The integration of the hedonic model with Geographical Information Systems (GIS) has expanded the ability of tourism researchers to value localized environmental amenities, such as green spaces, walkability, views, and noise levels \citep{lake2000using, bastian2002environmental, cavailhes2009gis, czembrowski2016integrating}. In the tourism literature, environmental quality is increasingly recognized as a key driver of destination competitiveness and hotel pricing.

A classic example is the valuation of scenic views, such as a sea view, which consistently commands a premium in coastal destinations \citep{fleischer2012room, latinopoulos2018using}. In urban tourism settings, the presence of green infrastructure (parks, vegetation density) represents a ``green premium'' by offering restorative environments that mitigate urban stressors like noise, congestion, and heat island effects \citep{wang2022contribution}. Using remote sensing indexes like the Normalized Difference Vegetation Index (NDVI) enables researchers to measure these environmental attributes objectively. However, the value of greenery is context-dependent: in dense urban centers, greenery is a scarce luxury that increases hotel value, whereas in peripheral, less-developed areas, dense vegetation may indicate a lack of urban development and transit infrastructure, leading to a pricing discount.

\begin{landscape}
\tiny
\setlength\LTleft{0pt}
\setlength\LTright{0pt}
\begin{longtable}{@{} >{\raggedright\arraybackslash}p{3.2cm} >{\raggedright\arraybackslash}p{3cm} >{\centering\arraybackslash}p{1cm} >{\raggedright\arraybackslash}p{5.0cm} p{9cm} @{}}
\caption{Comparative Analysis of Hotel Pricing Studies}
\label{tab:hotel_pricing_studies}\\
\toprule
\textbf{Authors} & \textbf{Research Area} & \textbf{Sample Size} & \textbf{Model Used} & \textbf{Independent Variables} \\
\midrule
\endfirsthead
\toprule
\textbf{Authors} & \textbf{Research Area} & \textbf{Sample Size} & \textbf{Model Used} & \textbf{Independent Variables} \\
\midrule
\endhead
\midrule 
\endfoot
\bottomrule
\endlastfoot
\citep{zhang2011modeling} & Beijing (China) & 228 & GWR & Hotel Characteristics (number of rooms,rating, year built), Location Attributes (distance) \\
\citep{zhang2011determinants} & New York (US) & 243 & OLS & Hotel Characteristics, Location Attributes, Review Attributes \\
\citep{lee2011room} & 6 US cities & 106 & OLS (Spatial Hedonic Price Model) & Location Attributes, Hotel Characteristics (amenities, type), Market Indicators \\
\citep{lee2012premium} & Downtown Chicago (US) & 81 & WLS, Spatial Lag Model & Location Attributes, Hotel Characteristics (size, amenity, affiliation) \\
\citep{santana2015influence} & La Palma (Spain), PengHu (Taiwan) & 190 & Spatial Econometric Model & Structural Features, Environmental Features, Location Attributes (remoteness, view-related isolation) \\
\citep{ert2016trust} & Stockholm (Sweden) & 395 & OLS & Trust-related Attributes, Review/Score-related Attributes, Host Characteristics, Apartment Characteristics \\
\citep{chen2017consumer} & Austin, Texas (US) & 5779 & OLS & Room Characteristics, Host Attributes, Cancellation Policy, Reputation, Market Competition \\
\citep{pawlicz2017determinants} & Warsaw (Poland) & 79 & Semi-Log Hedonic Model & Hotel Characteristics (size, quality indicators, affiliation), Location Attributes \\
\citep{latinopoulos2018using} & Halkidiki (Greece) & 557 & SGWR & Hotel/Room Characteristics, Environmental Attributes, Location Attributes \\
\citep{soler2018hedonic} & Malaga (Spain) & 57 & GWR & Hotel Characteristics (category, size), Location Attributes (distance to city center, train station) \\
\citep{tang2019estimating} & US cities & 654 & OLS, Spatial Error Model, Spatial Lag Model & Room Characteristics (bedrooms, capacity), Review Score, Location Indicators \\
\citep{soler2019algarve} & Algarve (Portugal) & 9992 & OLS & Hotel Characteristics (style, amenities), Reputational Attributes, Location Attributes, Contextual Attributes \\
\citep{sanchez2019modeling} & France, Spain, Italy, UK & 3800 & Quantile Regression & Digital Influence, Market Conditions, Hotel Attributes (star rating) \\
\citep{kim2020hotel} & Chicago (US) & 387 & GWR & Hotel Characteristics (size, age, class, service quality), Location Attributes (distance to airports, highways, attractions) \\
\citep{conroy2020effect} & Las Vegas (US) & 6087 & Semi-log Linear Regression (with Splines) & Room Characteristics, Hotel Characteristics, Time Variables, Location Attributes \\
\citep{hong2020analyzing} & Los Angeles and New York (US) & 44291 & MGWR & Listing functions, Attributes of host, Reputation, Geographical characteristics \\
\citep{bhattacharya2021spatial} & Pacific Coast (Japan) & 2813 & SGWR & Hotel Attributes, Environmental Attributes, Location Attributes, Regional Attributes \\
\citep{wang2022contribution} & Amsterdam (Netherlands) & 24150 & GWR & Reputational Attributes, Location Attributes \\
\citep{zhao2023multiscale} & Beijing (China) & 1499 & MGWR & Functional Attributes, Location Attributes, Reputation Attributes, Host Status \\
\citep{gordan2024hedonic} & Romania & 5028 & GWR & Accommodation Characteristics (size, capacity, facilities), Environmental Attributes \\
\citep{yang2025study} & Suzhou (China) & 81 & Exponential Linear Model, Semi-Log Hedonic Model & Architectural Features, Location Attributes, Business Attributes, Traffic Attributes, Facility Attributes \\
\end{longtable}
\end{landscape}

\subsection{Methodological Evolution and Research Gaps}

Reflecting the complexity of spatial data, a variety of sophisticated methodologies have been developed and employed in spatial hedonic analysis, spanning classic spatial econometrics, machine learning, and hybrid structures. Standard Ordinary Least Squares (OLS) models assume spatial independence, which is often violated in geographic data. To address spatial dependence, classic spatial econometric models like the Spatial Autoregressive (SAR) and Spatial Error Models (SEM) incorporate spatial weights matrices to capture spillovers and correlated error structures \citep{anselin1988spatial, seo2014combined}. Researchers continue to build upon this foundation with advanced techniques, such as M-quantile regression to address market heterogeneity \citep{schirripa2024spatial}, and parametric/non-parametric models incorporating spatial drift via penalized splines \citep{montero2018housing}. Furthermore, quantile regressions have been used to analyze the heterogeneous effects of price determinants across different hotel tiers, demonstrating that factors like online ratings have varying impacts over time and location \citep{wang2019tourism}.

More recently, there has been a trend towards integrating machine learning techniques to handle non-linear relationships and complex interactions that traditional models may miss. Algorithms like XGBoost have been combined with the hedonic framework to predict prices and analyze the comprehensive effects of influential factors \citep{zaki2022house, li2021understanding}. Similarly, a spatially explicit random forests approach has been introduced to inherently account for spatial autocorrelation in the data \citep{tepe2024random}, alongside bias-reduced spatial cross-validation strategies to improve model accuracy \citep{deppner2024accounting}. Hybrid and novel approaches increasingly combine multiple techniques to tackle specific challenges like endogeneity bias. For example, a regression spline model has been combined with machine learning, a novel matching algorithm, and a spatial difference-in-differences approach \citep{liu2024area}, while calibrated bid-rents from a spatial interaction model (SIM) are used as hedonic price estimators \citep{borba2016evaluation}. Among the most prominent methods specifically designed to address spatial heterogeneity is Geographically Weighted Regression (GWR) \citep{brunsdon1996geographically}. This powerful local regression technique estimates separate models for each observation, allowing the coefficients of the explanatory variables to vary across space, thereby explicitly mapping the heterogeneity in price determinants.

Despite these methodological advances, a surprising gap remains in the literature. First, standard hedonic or multiple linear regression models may fail to identify true causal effects if they neglect to account for unobserved hotel characteristics that are correlated with the variables of interest, such as a room's view \citep{dube2024hotel}. This highlights the need for careful model specification and accounting for spatial factors. Second, there has been a notable lack of comprehensive spatial hedonic modeling applied to emerging hospitality markets like India, which are characterized by rapid, multi-nodal development. Third, and most importantly, while spatial dependence and spatial heterogeneity are known to coexist due to localized competitive dynamics (coopetition) and environmental variation, studies that jointly estimate both within a single integrated framework remain extremely limited. This study fills these gaps by applying the integrated MGWR--SAR model proposed by \citet{geniaux2018new} to the budget hotel market of Delhi NCR, analyzing how localized price spillovers, transit mobilities, and green premiums vary across space.

\section{Methodology}
\label{sec:methodology}
\subsection{Study Area}

The Delhi National Capital Region (NCR) is a vast, interconnected metropolitan area in India that encompasses the National Capital Territory (NCT) of Delhi and several surrounding districts from the states of Haryana, Uttar Pradesh, and Rajasthan. Geographically, it stretches across a significant portion of northern India, forming a contiguous urban and semi-urban landscape.

Delhi, at its core, is a historically rich city, serving as India's capital and a major political, economic, and cultural center. The Delhi NCR's expansion includes key cities such as Gurugram (Haryana), Noida and Ghaziabad (Uttar Pradesh), and Faridabad (Haryana). The region is characterized by dynamic growth, diverse demographics, and strong infrastructure, including an extensive metro network, Indira Gandhi International Airport, and a rapidly expanding highway system that strengthens regional and inter-state connectivity.

According to the India Tourism Data Compendium \citep{indiatourism2024}, Delhi recorded 183,000 foreign tourist visits in 2023. Among major tourist attractions, Qutub Minar and the Red Fort witnessed the highest footfall from Indian visitors during the financial years 2022–23 and 2023–24, with visits increasing from 1.768 million to 3.124 million at Qutub Minar and from 2.215 million to 2.794 million at the Red Fort. Qutub Minar also attracted the highest number of foreign visitors during the same period, with footfall rising from 115,000 to 220,000. Air travel remained the preferred mode of entry for foreign tourists, accounting for 79.4\% of all Foreign Tourist Arrivals (FTAs), of which 57.2\% landed in Delhi. Delhi Airport continued to serve as a key international entry point, handling 3.074 million, 2.886 million, and 3.031 million FTAs in the years 2021, 2022, and 2023 respectively.

According to the Tourism Survey in the State of Delhi – Annual Final Report \citep{tourismdelhi}, accommodation preferences showed that 57\% of domestic tourists stayed with friends and relatives, while 41\% stayed in hotels. In contrast, 90\% of foreign tourists chose to stay in hotels. The Delhi NCR's rapid and uneven urbanization, together with its multi-nodal urban development pattern, makes the region an appropriate setting for examining spatial heterogeneity, green premiums, and spatial coopetition in accommodation pricing.

\subsection{Data Collection and Variables}   
The data for this study was collected from OYO Rooms (www.oyorooms.com) for the Delhi NCR region. The platform operates on a dynamic pricing model and provides a self-service interface to partner hotels, enabling them to adjust prices in response to demand conditions and optimize revenue. OYO primarily serves the budget accommodation segment, catering to both domestic and international travelers seeking affordable yet standardized lodging options. Focusing on this segment is especially relevant, as budget hotel customers tend to be highly price-sensitive, allowing for a clearer examination of how spatial, transport, and environmental factors influence pricing decisions.

\begin{figure}[H]
\centering
\begin{tikzpicture}

\node[anchor=south west,inner sep=0] (image) at (0,0)
{\includegraphics[width=0.9\linewidth]{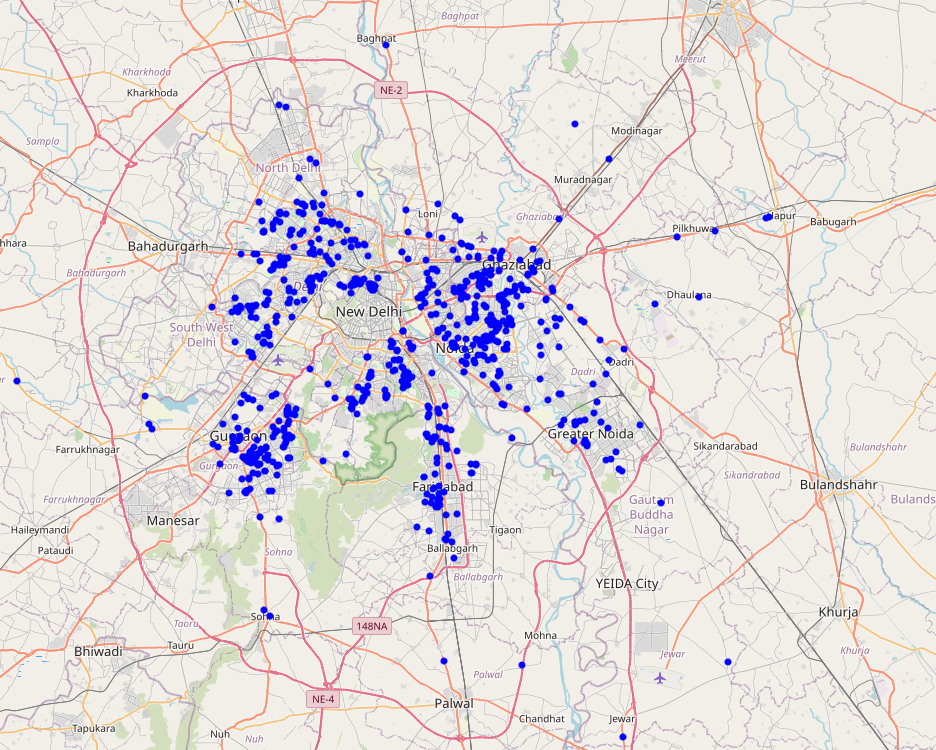}};

\draw[black, thick] (image.south west) rectangle (image.north east);

\begin{scope}[x={(image.south east)},y={(image.north west)}]
\node at (0.95,0.85) {\Large $\uparrow$};
\node at (0.95,0.80) {\textbf{N}};
\end{scope}

\end{tikzpicture}

\caption{Map showing the spatial distribution of OYO hotels within the Delhi NCR.}
\label{figure:hotels}

\end{figure}

The dataset includes room prices (inclusive of tax and after applying discounts) reserved for stays during the weekdays of July 2025 for a single day and for 2 adults, along with various attributes such as amenities, ratings, room availability, and location, for 857 hotels. In addition, the distance from each property to the nearest metro station, major tourist attraction, airport, and key bus stop was calculated to capture the influence of location accessibility and tourist mobilities on pricing.

Figure~\ref{figure:hotels} illustrates the spatial distribution of hotels across Delhi NCR. The distribution reveals a strong concentration of hotels within major urban centers, particularly in New Delhi, Gurugram, Noida, and Ghaziabad. Hotels are more densely clustered along key transportation corridors and commercial hubs, highlighting the influence of urban development patterns, transit accessibility, and local tourist agglomerations.

\begin{table}[htbp]
\centering
\caption{Summary of Data Sources Used in the Study}
\begin{tabular}{p{6cm} p{8cm}}
\hline
\textbf{Data Type / Variable} & \textbf{Source} \\
\hline
Hotel Listings & OYO Rooms website (\url{https://www.oyorooms.com}) \\[4pt]
Distance to City Center & OpenStreetMap (OSM) \\[4pt]
Distance from Nearest Metro Station & Delhi Metro Rail Corporation (DMRC) website \\[4pt]
Distance to Nearest Bus Stop & OpenStreetMap (OSM) \\[4pt]
Distance from Nearest Airport & OpenStreetMap (OSM) \\[4pt]
Distance to Tourist Attraction & OpenStreetMap (OSM) \\[4pt]
Built-up Land Cover & Harmonized Sentinel-2 MSI: MultiSpectral Instrument, Level-1C (TOA) \\[4pt]
Night-time Light Intensity & VIIRS Nighttime Day/Night Band Composites Version 1 \\[4pt]
Vegetation Density (NDVI) & Harmonized Sentinel-2 MSI: MultiSpectral Instrument, Level-2A (Surface Reflectance) \\
\hline
\end{tabular}
\label{tab:data_sources}
\end{table}

The explanatory variables used in this study comprise structural, locational, and environmental attributes, sourced from multiple open datasets and remote sensing platforms. Table~\ref{tab:data_sources} provides a detailed summary of the data sources and variable construction.

Structural attributes were extracted from the OYO Rooms platform and capture hotel-specific characteristics, including discount percentage, total number of rooms, number of images, average rating, number of ratings, and the number of amenities. Locational attributes were derived from OpenStreetMap (OSM) and the Delhi Metro Rail Corporation (DMRC). These include distances to key urban infrastructure and points of interest, such as the city center, nearest metro station, bus stop, railway station, airport, and tourist attractions, representing transit accessibility and tourist mobilities.

To capture the local environmental context and evaluate the green premium, the study incorporates spatial variables derived from satellite imagery. Specifically, vegetation density (NDVI), Night-time Light Intensity (NTL), and Built-up Area (Dynamic World) were computed as annual composites for 2025 within a 500-meter buffer around each hotel. These variables provide consistent proxies for local greenery, economic activity, and urban development.

The Normalized Difference Vegetation Index (NDVI), used to evaluate the green tourism premium, was computed from Sentinel-2 surface reflectance data \citep{drusch2012sentinel} using the normalized difference between the near-infrared (NIR) and red spectral bands as:
\begin{equation}
NDVI = \frac{B8 - B4}{B8 + B4},
\end{equation}
where $B8$ represents the near-infrared band (Sentinel-2 Band~8) and $B4$ represents the red band (Sentinel-2 Band~4). The final NDVI values were obtained by averaging cloud-filtered observations for July 2025.

Night-time Light Intensity was derived from the VIIRS Day/Night Band monthly composite data \citep{elvidge2017viirs} using the average radiance band (avg\_rad) as:
\begin{equation}
NTL_{\text{avg}} = \frac{1}{n} \sum_{i=1}^{n} L_{\text{cloud-free},i},
\end{equation}
where $L_{\text{cloud-free},i}$ denotes the radiance observed during the $i^{\text{th}}$ cloud-free night, and $n$ is the total number of observations.

The Built-up Area was derived from Dynamic World data \citep{brown2022dynamic}, which uses a fully convolutional neural network (FCNN) applied to Sentinel-2 imagery to estimate land-cover probabilities:
\begin{equation}
P(\text{Built}) = \text{FCNN}\left(S2_{\text{Bands}},\, S2_{\text{Spatial}}\right),
\end{equation}
where $S2_{\text{Bands}}$ represents the multispectral Sentinel-2 inputs, and $S2_{\text{Spatial}}$ denotes the surrounding spatial context. Built-up Area ($A_{\text{built}}$) within a region of interest was calculated by aggregating pixel-level classifications:
\begin{equation}
A_{\text{built}} = \sum_{i=1}^{n} \left( \text{Mask}_{\text{built},i} \times \text{Area}_{\text{pixel},i} \right),
\end{equation}
where $\text{Mask}_{\text{built},i}$ is a binary indicator for built-up classification, and $\text{Area}_{\text{pixel},i}$ is the physical area of a single pixel ($100\,\text{m}^2$).

\begin{figure}[p]
\centering

\makebox[\textwidth]{\makebox[0.95\textwidth][l]{\textbf{(a)}}}
\includegraphics[height=0.255\textheight,keepaspectratio]{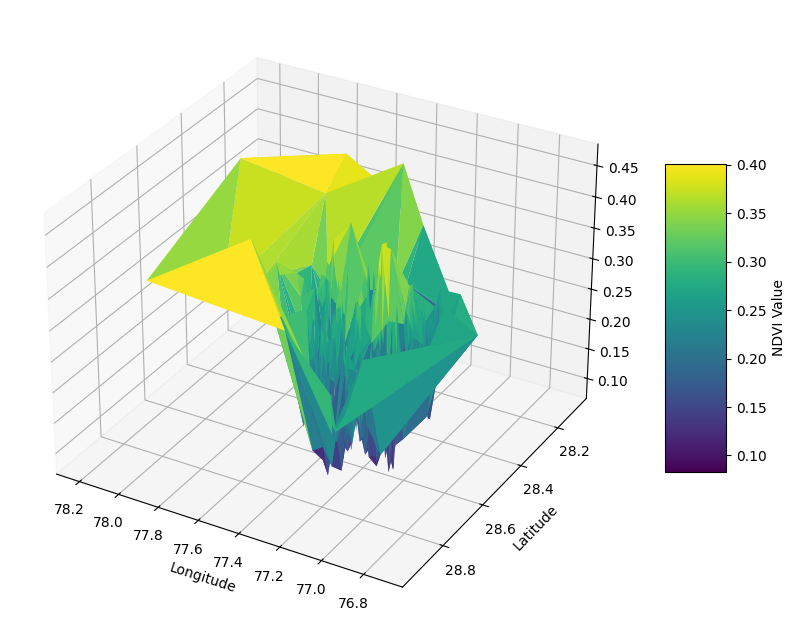}

\vspace{0.25cm}

\makebox[\textwidth]{\makebox[0.95\textwidth][l]{\textbf{(b)}}}
\includegraphics[height=0.255\textheight,keepaspectratio]{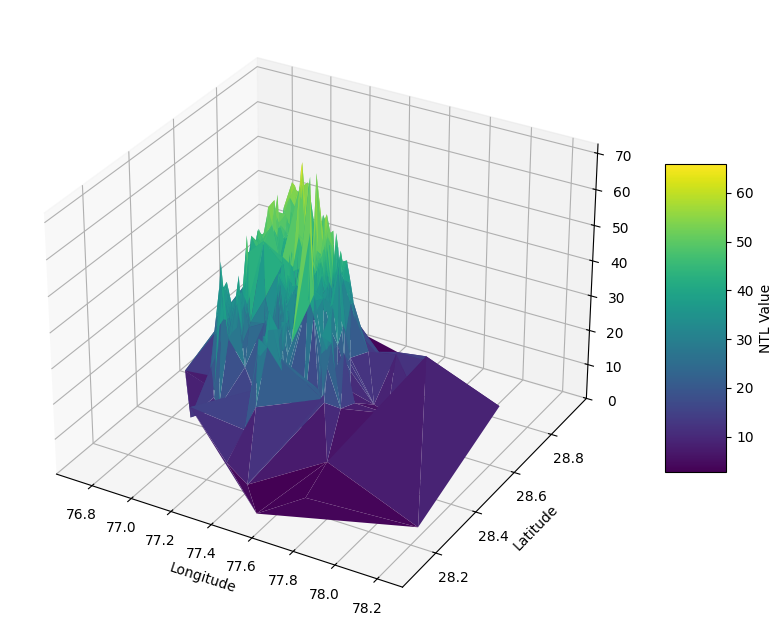}

\vspace{0.25cm}

\makebox[\textwidth]{\makebox[0.95\textwidth][l]{\textbf{(c)}}}
\includegraphics[height=0.255\textheight,keepaspectratio]{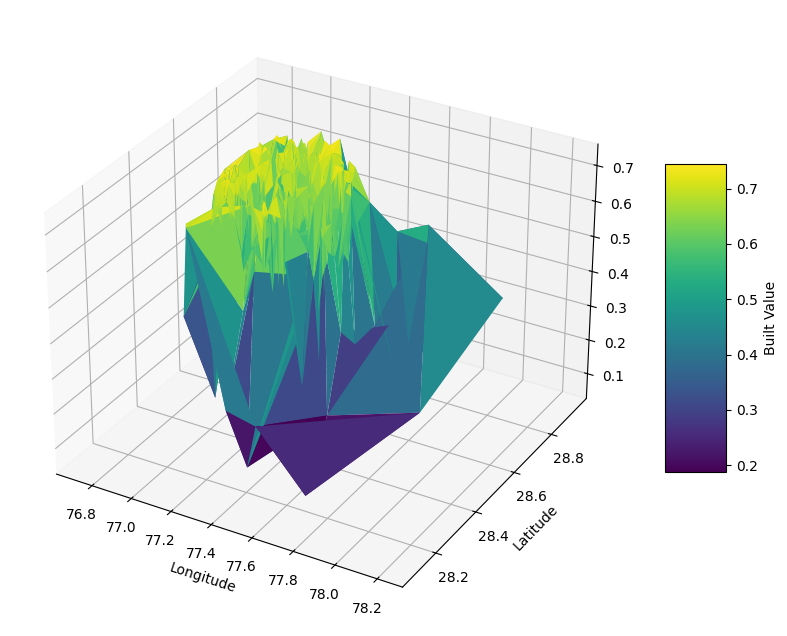}

\caption{Three-dimensional surface plots illustrating spatial variation in environmental indicators across the study area: (a) NDVI representing vegetation intensity, (b) Night-time Light Intensity indicating urban activity, and (c) Built-up Area showing the distribution of urbanized regions.}

\label{figure:combined_surface}
\end{figure}

\begin{figure}[H]
    \centering
    \includegraphics[width=1 \linewidth]{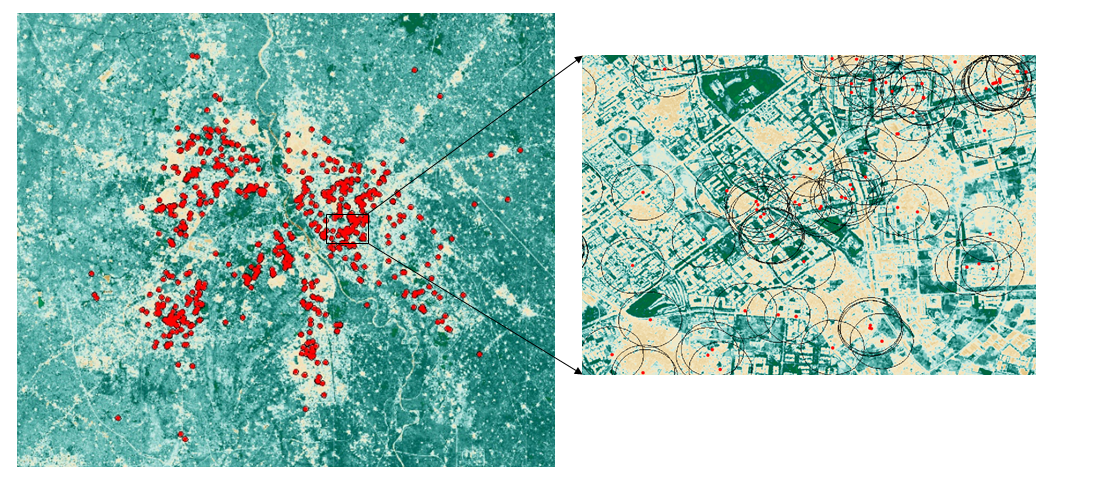}
    \caption{Buffer region for vegetation index}
    \label{figure:500mbuffer}
\end{figure}

The integration of platform-based data and satellite-derived indicators enables a multidimensional characterization of hotel environments, improving the ability to capture environmental quality and economic context. Figure~\ref{figure:combined_surface} presents the environmental surfaces across the region. Complementing these surfaces, Figure~\ref{figure:500mbuffer} illustrates the vegetation density within a 500-meter buffer around each hotel.

\subsection{Hedonic Modelling}

\begin{figure}[htbp]
    \centering
    \includegraphics[width=\textwidth]{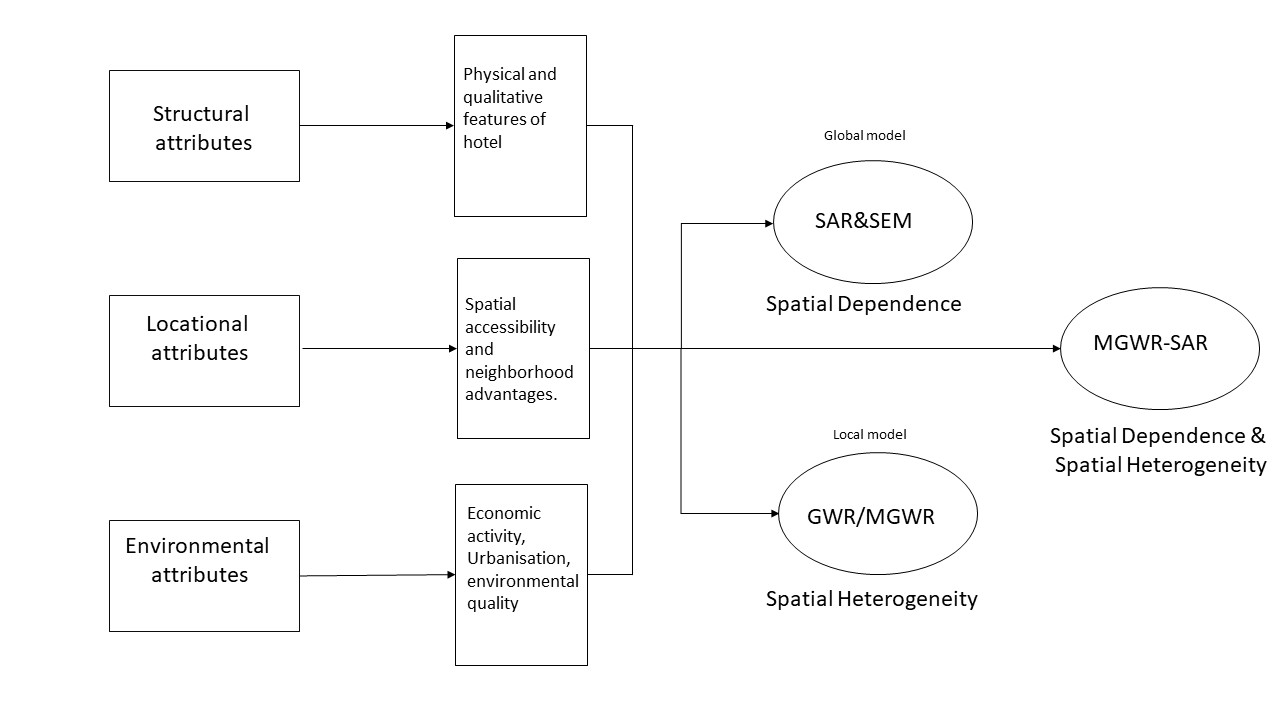}
    \caption{Modelling Framework}
    \label{figure:maps}
\end{figure}

Following the seminal contributions of \citep{lancaster1966new} and \citep{rosen1974hedonic}, the price of a hotel room is modeled as a function of its structural, locational, and environmental attributes. Let $y_i = \ln(P_i)$ denote the log price of hotel room $i$. Formally, the hedonic price function is expressed as:
\begin{equation}
\ln P_i = f(\mathbf{S}_i, \mathbf{L}_i, \mathbf{E}_i),
\end{equation}
where $P_i$ denotes the price of hotel room $i$, $\mathbf{S}_i = (S_{i1}, \dots, S_{iK})$ is a vector of structural characteristics, $\mathbf{L}_i = (L_{i1}, \dots, L_{iM})$ is a vector of transit locational attributes, and $\mathbf{E}_i = (E_{i1}, \dots, E_{iR})$ is a vector of environmental factors. Figure~\ref{figure:maps} presents the overall modelling framework.

The analysis begins with a baseline Ordinary Least Squares (OLS) model to establish global relationships. To account for spatial dependence, we extend the analysis using SAR and SEM models, which capture spatial spillovers and spatial error correlations. Subsequently, to address spatial heterogeneity, we employ GWR, allowing coefficients to vary across locations. Finally, we implement the integrated MGWR–SAR model, capturing both coopetition spillovers and environmental spatial heterogeneity simultaneously.

\subsubsection{Baseline OLS Model}
The benchmark OLS model assumes spatial independence and parameter homogeneity:
\begin{equation}
\label{eq:ols}
y_i = \beta_0 + \sum_{k=1}^{K} \beta_k X_{i,k} + \varepsilon_i,
\end{equation}
where \( y_i \) is the log price, \( X_{i,k} \) is the \( k^{\text{th}} \) explanatory variable, \( \beta_k \) are parameters to be estimated, and \( \varepsilon_i \) is an iid error term.

\subsubsection{Spatial Weights Matrix Construction}
To model spatial spillovers and competitive interactions, a spatial weights matrix \( W \) is constructed. Locations were geocoded and reprojected into the UTM Zone 43N (EPSG:32643) projected coordinate system. A $k$-nearest-neighbor (kNN) approach is employed with $k=4$, representing the closest competing properties. The neighbor structure is row-standardized (style ``W''), ensuring that the weights for each observation sum to one: $\sum_{j} w_{ij} = 1$ for all $i.$ The weights are constructed in R using the sequence: \texttt{knearneigh} $\rightarrow$ \texttt{knn2nb} $\rightarrow$ \texttt{nb2listw(style = "W",\, zero.policy = TRUE)}.

\subsubsection{Spatial Diagnostics}
The Global Moran’s I statistic is computed on the OLS residuals to test for spatial autocorrelation:
\begin{equation}
I = \frac{n}{S_0} \cdot 
\frac{\sum_{i}\sum_{j} w_{ij}(y_i - \bar{y})(y_j - \bar{y})}
{\sum_{i}(y_i - \bar{y})^2},
\end{equation}
where \( n \) is the number of observations, \( y_i \) and \( y_j \) represent log prices, \( \bar{y} \) is the sample mean, \( w_{ij} \) are elements of the spatial weight matrix \( W \), and \( S_0 = \sum_i \sum_j w_{ij} \) is the sum of all spatial weights. A significant positive value indicates pricing clustering, while a negative value suggests spatial dispersion.

\subsubsection{Spatial Dependence Models - SAR and SEM}
The SAR model incorporates a spatially lagged dependent variable to capture direct price spillovers and competition:
\begin{equation}
\label{eq:sar}
y_i = \rho \sum_{j=1}^{n} w_{ij} y_j + \beta_0 + \sum_{k=1}^{K} \beta_k X_{i,k} + \varepsilon_i,
\end{equation}
where \( \rho \) is the global spatial autoregressive coefficient, and \( \varepsilon_i \) is an iid disturbance term.

Alternatively, the SEM captures spatial autocorrelation in unobserved factors through the error term:
\begin{align}
y_i &= \beta_0 + \sum_{k=1}^{K} \beta_k X_{i,k} + u_i, \\
u_i &= \lambda \sum_{j=1}^{n} w_{ij} u_j + \varepsilon_i,
\label{eq:sem}
\end{align}
where \( \lambda \) measures spatial autocorrelation in the error process, and \( \varepsilon_i \) is a white-noise error term.

\subsubsection{Spatial Heterogeneity Model - GWR}
To capture spatial heterogeneity, GWR allows regression coefficients to vary continuously across coordinates:
\begin{equation}
y_i = \sum_{k=0}^{p} \beta_k(u_i, v_i)\, x_{ik} + \varepsilon_i,
\end{equation}
where $(u_i, v_i)$ are the spatial coordinates of observation $i$, and $\beta_k(u_i, v_i)$ are location-specific parameter estimates. Local parameters are estimated using weighted least squares:
\begin{equation}
\hat{\boldsymbol{\beta}}(u_i, v_i) = \left(X^{T} W(u_i, v_i) X\right)^{-1} X^{T} W(u_i, v_i) y,
\end{equation}
where $W(u_i, v_i)$ is a spatial kernel matrix. The kernel bandwidth is selected by minimizing the corrected Akaike Information Criterion (AICc).

\subsection{Integrated Model - MGWR--SAR}
To jointly account for spatial heterogeneity and spatial dependence (local pricing dynamics), the analysis employs the MGWR--SAR model proposed by \citet{geniaux2018new}:
\begin{equation}
y_i = \rho(u_i, v_i)\,(Wy)_i + \sum_{k=1}^{k_c} \beta_k x_{ik} + \sum_{k=k_c+1}^{p} \beta_k(u_i, v_i)\, x_{ik} + \varepsilon_i,
\end{equation}
where $(Wy)_i$ is the spatial lag of log price, $\rho(u_i, v_i)$ is the spatially varying spatial autoregressive parameter, $k_c$ is the number of spatially constant coefficients, and $k_v = p - k_c$ is the number of spatially varying coefficients. 

In this study, we estimate the MGWR--SAR$(1,0,k_v)$ specification, where the spatial autoregressive parameter and all regression coefficients vary across space, allowing full local flexibility in both spillover intensity and covariate effects:
\begin{equation}
y_i = \rho(u_i, v_i)\,(Wy)_i + \sum_{k=0}^{p} \beta_k(u_i, v_i)\, x_{ik} + \varepsilon_i.
\end{equation}

Because the spatial lag term $(Wy)_i$ is endogenous, estimation is performed using Spatial Two-Stage Least Squares (S2SLS) with the local instrument matrix:
\begin{equation}
H = [X, WX, W^2X, W^3X],
\end{equation}
yielding the estimator of local parameters $\theta(u_i,v_i)$ as:
\begin{equation}
\hat{\boldsymbol{\theta}} = \left(X^{\top} P_H X\right)^{-1} X^{\top} P_H y, \quad P_H = H(H^{\top}H)^{-1}H^{\top}.
\end{equation}

Spatial heterogeneity is captured using an adaptive Gaussian kernel, and the optimal bandwidth is selected by minimizing the corrected Akaike Information Criterion (AICc) using a multi-round grid search with golden-section refinement.

Because of spatial feedback loop spillovers, MGWR--SAR coefficients reflect both direct and indirect (spillover) effects. Hence, this study focuses on identifying regions where spatial dependence is statistically significant and examining the geographic variation in local coefficients. All statistical computations are conducted in R (version 4.5.1) and Python (3.13.7).

\section{Results}

This section provides estimates obtained from global (OLS, SAR, and SEM) and local (MGWR--SAR) models and highlights differences arising from spatial heterogeneity and dependence. Table~\ref{tab:summary_vif} presents descriptive statistics and Variance Inflation Factors (VIFs) for the explanatory variables. Hotel prices range from 486 to 3,971, with a mean of 1,123 and a median of 1,056, indicating a slight right-skewness in the distribution. All VIF values remain below the commonly accepted threshold of 5, suggesting the absence of serious multicollinearity and ensuring the stability of coefficient estimates across models.

\begin{table}[H]
\centering
\caption{Summary Statistics and Variance Inflation Factors (VIF) for Explanatory Variables}
\label{tab:summary_vif}
\begin{threeparttable}
\resizebox{\textwidth}{!}{%
\begin{tabular}{lccccccc}
\toprule
\textbf{Variable} & \textbf{Min} & \textbf{1st Quartile} & \textbf{Median} & \textbf{Mean} & \textbf{3rd Quartile} & \textbf{Max} & \textbf{VIF} \\
\midrule
Price                                 &  486  &   859 &   1056 &   1123 &   1231 &   3971 & -  \\
Discount Percentage (\%)              & 41.80 & 75.50 & 77.70 & 76.92 & 78.00 & 86.80 & 1.11 \\
Total Available Rooms                 & 0.00  & 11.00 & 18.00 & 20.98 & 27.00 & 154.00 & 1.14 \\
Number of Images                      & 4.00  & 27.00 & 30.00 & 32.33 & 34.00 & 118.00 & 1.07 \\
Average Rating                        & 2.00  & 3.70  & 4.11  & 4.11  & 4.60  & 5.00  & 1.09 \\
Number of Ratings                     & 0.00  & 8.00  & 47.00 & 245.20 & 244.30 & 5223.00 & 1.20 \\
Number of Amenities                   & 1.00  & 5.00  & 7.00  & 9.13  & 12.00  & 38.00  & 1.14 \\
Distance to City Center (km)          & 1.05  & 12.29 & 16.35 & 17.96 & 23.30 & 68.02 & 2.71 \\
Distance to Metro (km)                & 0.01  & 0.62  & 1.43  & 2.36  & 2.76  & 34.52 & 2.30 \\
Distance to Railway Station (km)      & 0.34  & 4.07  & 6.69  & 7.32  & 9.12  & 36.94 & 2.70 \\
Distance to Bus Stop (km)             & 0.12  & 2.78  & 4.46  & 5.54  & 6.97  & 40.21 & 3.33 \\
Distance to Airport (km)              & 3.08  & 12.16 & 14.91 & 16.57 & 19.59 & 70.13 & 1.82 \\
Distance to Tourist Attraction (km)   & 0.35  & 2.68  & 6.62  & 7.84  & 11.27 & 54.52 & 2.80 \\
Built-up Area (normalized)            & 0.04  & 0.63  & 0.68  & 0.66  & 0.71  & 0.74  & 2.47 \\
Night-time Light Intensity           & 1.34  & 39.48 & 47.31 & 48.00 & 55.69 & 120.72 & 1.49 \\
NDVI                                  & 0.07  & 0.18  & 0.22  & 0.22  & 0.26  & 0.48  & 2.03 \\
\bottomrule
\end{tabular}%
}
\end{threeparttable}
\end{table}

\subsection{Results from OLS, SAR and SEM Models}

Table~\ref{tab:ols_sar_sem_full} reports the global OLS, SAR, and SEM estimates. A Global Moran's $I$ test on OLS residuals indicates significant positive spatial autocorrelation (Moran's $I = 0.079$, $p < 0.001$). Coefficient signs and magnitudes remain broadly stable across specifications, suggesting robust relationships between hotel prices and their structural, locational, and environmental attributes. The spatial models improve fit relative to OLS, and the significant spatial parameters confirm the presence of spatial dependence.

Rao's score (Lagrange Multiplier) diagnostics also support spatial dependence in the OLS residuals. Both the LM error test (RSerr = 12.552, $p < 0.001$) and LM lag test (RSlag = 11.953, $p < 0.001$) are significant, although the robust versions are not. This indicates that spatial effects are present, but neither the lag nor error specification clearly dominates. The estimated spatial parameters are positive and significant in both models. The SAR model yields $\rho = 0.139$, consistent with positive price spillovers across nearby hotels, while the SEM model reports $\lambda = 0.165$, suggesting that unobserved spatially correlated factors such as neighborhood quality or local regulation also matter.

Model diagnostics reinforce this conclusion. Both SAR and SEM outperform OLS in terms of AIC and residual variance, with SEM providing the best overall fit. The OLS model explains about $34.2\%$ of the variation in hotel prices, based on an adjusted $R^2$ of $0.330$. 

Across all global models, discount percentage emerges as the most influential determinant, exhibiting a large and consistently negative effect on prices ($\approx -3.15$), underscoring its central role in pricing strategy. Among structural attributes, hotel size (Total Available Rooms) and Built-up Area have positive and statistically significant effects, although the magnitude of the Built-up Area coefficient declines slightly in spatial models, suggesting that part of its effect is spatially structured.

Reputation-related variables further highlight the importance of perceived quality. Average Rating shows a strong and robust positive association with price, whereas the Number of Ratings remains insignificant, indicating that quality signals matter more than sheer review volume. The Number of Amenities contributes a modest but positive premium.

Locational variables reveal clear spatial gradients. Greater distance from the city center and metro stations is associated with lower prices, reflecting reduced accessibility, while the global effect of distance to railway stations is negative but not statistically significant. In contrast, distance to bus stops shows a positive relationship, potentially capturing quieter or more premium neighborhoods. The effects of distance to tourist attractions and distance to the airport are not statistically significant in the global models.

Finally, environmental factors also play a role. Vegetation density (NDVI) consistently exerts a positive effect on prices, suggesting a premium for greener surroundings. In contrast, Night-time Light Intensity does not have a significant independent effect, indicating that once other locational and structural factors are accounted for, its contribution is limited.

\begin{table}[H]
\centering
\caption{Coefficient Estimates: OLS, Spatial Lag (SAR – Direct Effects) and Spatial Error (SEM) Models}
\label{tab:ols_sar_sem_full}
\footnotesize
\setlength{\tabcolsep}{4pt}
\begin{threeparttable}
\begin{tabular}{lccc}
\toprule
\textbf{Variable} & \textbf{OLS} & \textbf{SAR (Direct)} & \textbf{SEM (error)} \\
\midrule
Intercept & $9.031^{***}$ & $8.090^{***}$ & $9.029^{***}$ \\
Discount Percentage & $-3.158^{***}$ & $-3.166^{***}$ & $-3.136^{***}$ \\
Total Available Rooms & $0.239^{**}$ & $0.251^{**}$ & $0.247^{**}$ \\
Number of Images & $0.101$ & $0.101$ & $0.099$ \\
Average Rating & $0.175^{***}$ & $0.179^{***}$ & $0.181^{***}$ \\
Number of Ratings & $-0.163$ & $-0.163$ & $-0.157$ \\
Number of Amenities & $0.112$ & $0.113^{*}$ & $0.114^{*}$ \\
Distance to City Center (km) & $-0.319^{***}$ & $-0.276^{***}$ & $-0.314^{**}$ \\
Distance to Metro (km) & $-0.181^{**}$ & $-0.172^{**}$ & $-0.183^{*}$ \\
Distance to Railway Station (km) & $-0.115$ & $-0.132$ & $-0.120$ \\
Distance to Bus Stop (km) & $0.220^{**}$ & $0.212^{**}$ & $0.226^{**}$ \\
Distance to Airport (km) & $0.005$ & $0.011$ & $0.000$ \\
Distance to Tourist Attraction (km) & $0.085$ & $0.076$ & $0.082$ \\
Built-up Area & $0.321^{**}$ & $0.274^{**}$ & $0.291^{**}$ \\
Night-time Light Intensity & $-0.109$ & $-0.092$ & $-0.075$ \\
NDVI & $0.351^{***}$ & $0.318^{***}$ & $0.337^{***}$ \\
\midrule
Spatial parameter & \multicolumn{2}{c}{ } & \\
\quad Spatial lag ($\rho$) & \multicolumn{1}{c}{---} & $0.139^{**}$ & --- \\
\quad Spatial error ($\lambda$) & \multicolumn{1}{c}{---} & --- & $0.165^{***}$ \\
\midrule
\textbf{Diagnostics} & & & \\
No.\ observations & \multicolumn{1}{c}{857} & \multicolumn{1}{c}{857} & \multicolumn{1}{c}{857} \\
AIC & 52.94 & 44.68 & 43.54 \\
Residual variance ($\hat{\sigma}^2$) & 0.0610 & 0.0589 & 0.0588 \\
OLS $R^2$ / Adj.\ $R^2$ & 0.342 / 0.330 & \multicolumn{1}{c}{--} & \multicolumn{1}{c}{--} \\
Log-likelihood & \multicolumn{1}{c}{--} & -4.339 & -3.770 \\
ML / number of params & \multicolumn{1}{c}{15 \& 841 DF} & 18 params & 18 params \\
\bottomrule
\end{tabular}
\end{threeparttable}
\end{table}

\subsection{Results from MGWR--SAR Model}
\begin{figure}[!htbp]
    \centering
    \makebox[\textwidth]{%
        \makebox[0.5\textwidth][c]{\textbf{(a)}}%
        \makebox[0.5\textwidth][c]{\textbf{(b)}}%
    }
    \vspace{0.5em}
    \includegraphics[width=\textwidth,height=0.6\textheight,keepaspectratio]{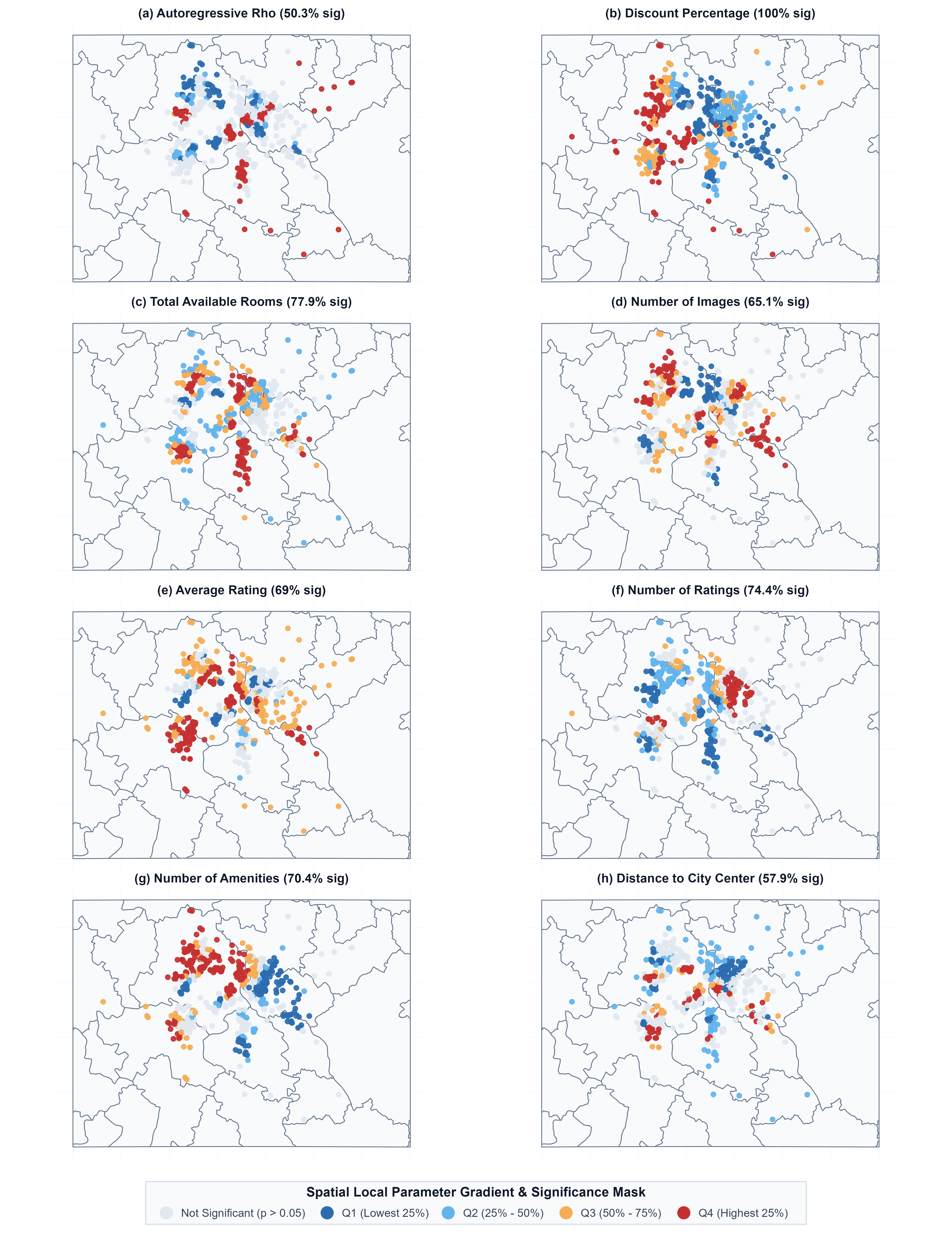}
    \caption{MGWR--SAR local spatial autoregressive parameter and significance maps.}
    \label{fig:mgwrsar_lambda}
\end{figure}

The MGWR--SAR$(1,0,k_v)$ estimates reported in Table~\ref{tab:mgwrsar_detailed} support the hypothesis that spatial interactions are not constant over space, as both the sign and magnitude of the spatial dependence parameter vary across sub-markets. The distribution of MGWR--SAR coefficients (quantiles and minimum--maximum values) is illustrated in Figures~\ref{fig:mgwrsar_lambda}--\ref{fig:mgwrsar_environmental}; the corresponding numerical summary is provided in Table~\ref{tab:mgwrsar_detailed}. In addition, the model achieves a lower RMSE relative to competing specifications, indicating improved predictive performance. Bootstrap tests (Table~\ref{tab:Bootstrap_test}) provide strong statistical evidence that each covariate exhibits a significant spatially varying effect within the MGWR--SAR framework. Accordingly, the null hypothesis $H_0: \beta_k(u_i, v_i) = 0$ is rejected for all explanatory variables ($p < 0.01$). Further, the local spatial autoregressive parameter $\rho(u_i,v_i)$ is found to be statistically significant in 444 out of 857 regions (51.81\%), indicating spatial dependence is a key component of the data-generating process for approximately half the study area. In regions where $\rho(u_i,v_i)$ is not significant, local price formation appears to be driven more by structural and locational attributes, rather than systematic spatial spillover effects.

\begin{table}[ht]
\centering
\caption{Summary Statistics of Local Coefficients and Percentage of Significant Regions (MGWR--SAR$(1,0,k_v)$)}
\label{tab:mgwrsar_detailed}
\resizebox{\textwidth}{!}{
\begin{tabular}{lrrrrrrr}
\hline
Variable & Min & Q1 & Median & Mean & Q3 & Max & \% Significant \\
\hline
Intercept & 3.086 & 8.287 & 9.224 & 9.492 & 10.158 & 20.235 & 97.55 \\
Discount Percentage & -6.523 & -4.739 & -4.216 & -3.989 & -3.169 & -1.412 & 100.00 \\
Total Available Rooms & -0.115 & 0.149 & 0.267 & 0.300 & 0.369 & 1.205 & 73.98 \\
Number of Images & -0.979 & -0.028 & 0.101 & 0.062 & 0.181 & 0.611 & 47.26 \\
Average Rating & -0.130 & 0.047 & 0.122 & 0.130 & 0.216 & 0.375 & 66.04 \\
Number of Ratings & -1.084 & -0.490 & -0.247 & -0.231 & -0.043 & 0.559 & 66.74 \\
Number of Amenities & -0.298 & -0.079 & 0.046 & 0.110 & 0.218 & 1.242 & 57.41 \\
Distance to City Center & -9.393 & -0.304 & 0.069 & 0.580 & 0.707 & 20.860 & 31.97 \\
Distance to Metro & -5.089 & -0.339 & -0.161 & -0.264 & 0.065 & 0.895 & 54.96 \\
Distance to Railway Station & -2.466 & -0.574 & -0.237 & -0.251 & 0.033 & 1.893 & 50.64 \\
Distance to Bus Stop & -4.586 & -0.005 & 0.208 & 0.075 & 0.323 & 1.223 & 53.79 \\
Distance to Airport & -13.013 & -0.323 & -0.022 & -0.272 & 0.182 & 5.609 & 29.52 \\
Distance to Tourist Attraction & -3.285 & -0.217 & 0.084 & 0.180 & 0.327 & 4.393 & 40.72 \\
Built-up Area & -1.046 & 0.092 & 0.442 & 0.345 & 0.624 & 1.302 & 72.81 \\
Night-time Light Intensity & -0.926 & -0.191 & -0.073 & -0.054 & 0.110 & 0.503 & 38.04 \\
NDVI & -1.403 & 0.181 & 0.338 & 0.247 & 0.488 & 0.884 & 86.23 \\
$\rho(u_i,v_i)$ & -0.650 & -0.163 & 0.049 & -0.010 & 0.183 & 0.403 & 51.81 \\
\hline
\end{tabular}
}
\end{table}

\begin{table}[ht]
\centering
\caption{Bootstrap Test of Joint Significance of Spatially Varying Coefficients}
\label{tab:Bootstrap_test}
\begin{tabular}{lcc}
\hline
Variable & Bootstrap Statistic ($T$) & Bootstrap $p$-value \\
\hline
Discount Percentage & 13.360 & $< 0.01$ \\
Total Available Rooms & 3.547 & $< 0.01$ \\
Number of Images & 3.333 & $< 0.01$ \\
Average Rating & 3.913 & $< 0.01$ \\
Number of Ratings & 3.405 & $< 0.01$ \\
Number of Amenities & 3.469 & $< 0.01$ \\
Distance to City Center & 3.823 & $< 0.01$ \\
Distance to Metro & 3.409 & $< 0.01$ \\
Distance to Railway Station & 3.527 & $< 0.01$ \\
Distance to Bus Stop & 3.702 & $< 0.01$ \\
Distance to Airport & 3.315 & $< 0.01$ \\
Distance to Tourist Attraction & 3.412 & $< 0.01$ \\
Built-up Area & 3.577 & $< 0.01$ \\
Night-time Light Intensity & 3.293 & $< 0.01$ \\
NDVI & 3.851 & $< 0.01$ \\
\hline
\end{tabular}
\end{table}

Figure~\ref{fig:mgwrsar_lambda}(a)--(b) illustrates the spatial distribution of the locally estimated spatial autoregressive parameter $\rho(u_i,v_i)$, along with the corresponding map of statistically significant estimates. The results reveal pronounced spatial heterogeneity in the nature of spatial interactions, with both the magnitude and sign of $\rho(u_i,v_i)$ varying across sub-markets.

Negative values of $\rho(u_i,v_i)$ are predominantly concentrated in western and north-western clusters. Such negative spatial dependence indicates competitive or substitution effects, where higher neighboring prices are associated with relatively lower local prices, conditional on other factors. In contrast, positive and relatively stronger values of $\rho(u_i,v_i)$ are observed in the central and eastern clusters, suggesting reinforcing spatial dependence. In these regions, higher neighboring prices are positively associated with local prices, consistent with agglomeration effects or neighborhood-based reinforcement dynamics. Overall, these patterns underscore that spatial interactions are context-specific. In some sub-markets, competitive effects dominate, while in others, clustering effects prevail. This further supports our choice of the MGWR--SAR model, which simultaneously captures both spatial dependence and spatial heterogeneity.  

The accompanying significance map indicates that statistically reliable spatial dependence is concentrated primarily in central and northern zones, while peripheral areas exhibit fewer significant estimates. This suggests that both the strength and statistical significance of spatial feedback effects vary systematically across space. Moreover, the presence of both positive and negative $\rho$ estimates confirms that spatial interaction mechanisms are not uniform but instead context-specific and regionally differentiated.

These results provide strong empirical support for the MGWR--SAR framework. A conventional global SAR model, which imposes a single constant spatial parameter, would fail to capture the localized spillover patterns observed here. The spatial variability of $\rho(u_i,v_i)$ highlights that spatial dependence itself is geographically heterogeneous, underscoring the importance of adopting a spatially adaptive econometric specification. We next turn to the interpretation of the remaining figures, focusing on how various attributes influence the outcomes across space. \\

\textbf{Structural Attributes}\\

Structural characteristics exhibit notable spatial variation in their effects. Discount Percentage, shown in Figure~\ref{fig:mgwrsar_structural_a}(a)--(b), displays a uniformly negative and statistically significant impact across all regions, with the magnitude of the effect being stronger in the eastern and north-eastern clusters. 

In contrast, Total Available Rooms, depicted in Figure~\ref{fig:mgwrsar_structural_a}(c)--(d), shows heterogeneous effects, with negative coefficients concentrated in eastern areas and positive effects observed in western and central zones. Approximately 73.98\% of the local estimates are statistically significant, indicating a strong but spatially varying relationship. 

Similarly, Number of Amenities, presented in Figure~\ref{fig:mgwrsar_structural_b}(k)--(l), exhibits spatially differentiated effects, with negative effects in eastern and south-eastern clusters and positive effects in western regions. Around 57.41\% of local coefficients are significant, further reinforcing the presence of spatial heterogeneity in structural attributes.

Reputation and visibility-related attributes also display pronounced spatial heterogeneity. Average Rating (Figure~\ref{fig:mgwrsar_structural_b}(g)--(h)), Number of Ratings (Figure~\ref{fig:mgwrsar_structural_b}(i)--(j)), and Number of Images (Figure~\ref{fig:mgwrsar_structural_a}(e)--(f)) show both positive and negative effects across the study area. For Average Rating, negative effects are concentrated in central and north-eastern clusters, while positive effects dominate western and southern areas, with about 66.04\% of estimates statistically significant. Number of Ratings follows a similar pattern, with 66.74\% significance, indicating geographically varying valuation of review volume. Number of Images also varies in sign, with about 47.26\% of coefficients significant, reflecting context-dependent effects of listing visibility.\\

\textbf{Locational Attributes}\\

Accessibility variables show clear spatial heterogeneity across the study area. Distance to City Center (Figure~\ref{fig:mgwrsar_locational_a}(a)--(b)) and Distance to Tourist Attraction (Figure~\ref{fig:mgwrsar_locational_b}(k)--(l)) exhibit a proximity premium in northern and central clusters (negative coefficients), while western and some peripheral areas show positive effects, indicating that centrality does not uniformly enhance value. Overall, with 31.97\% to 40.72\% of regions showing statistical significance, the results confirm that accessibility effects are spatially heterogeneous rather than uniform.\\

\textbf{Environmental Attributes} \\
\subparagraph{Spatial Heterogeneity in the Effect of Built-up Area :}  {\color{black}  

The coefficient surface in Figure~\ref{fig:mgwrsar_environmental}(a)--(b), demonstrates clear spatial heterogeneity, with predominantly positive effects across much of the study area. Strong positive coefficients in central and north-eastern clusters indicate a size premium, where larger properties command higher prices. In contrast, negative or weaker effects in western and peripheral regions suggest that the marginal value of Built-up Area is lower or even adverse in those locations, possibly reflecting differences in land-use intensity or local demand conditions. Approximately $72.81\%$ of local coefficients are statistically significant, with significance concentrated in central and north-eastern zones, while peripheral areas show more uncertainty.  

\clearpage
\thispagestyle{empty}
\begin{figure}[!htbp]
    \centering
    \makebox[\textwidth]{%
        \makebox[0.5\textwidth][c]{\textbf{(a)}}%
        \makebox[0.5\textwidth][c]{\textbf{(b)}}%
    }
    \vspace{0.5em}
    \includegraphics[width=\textwidth,height=0.6\textheight,keepaspectratio]{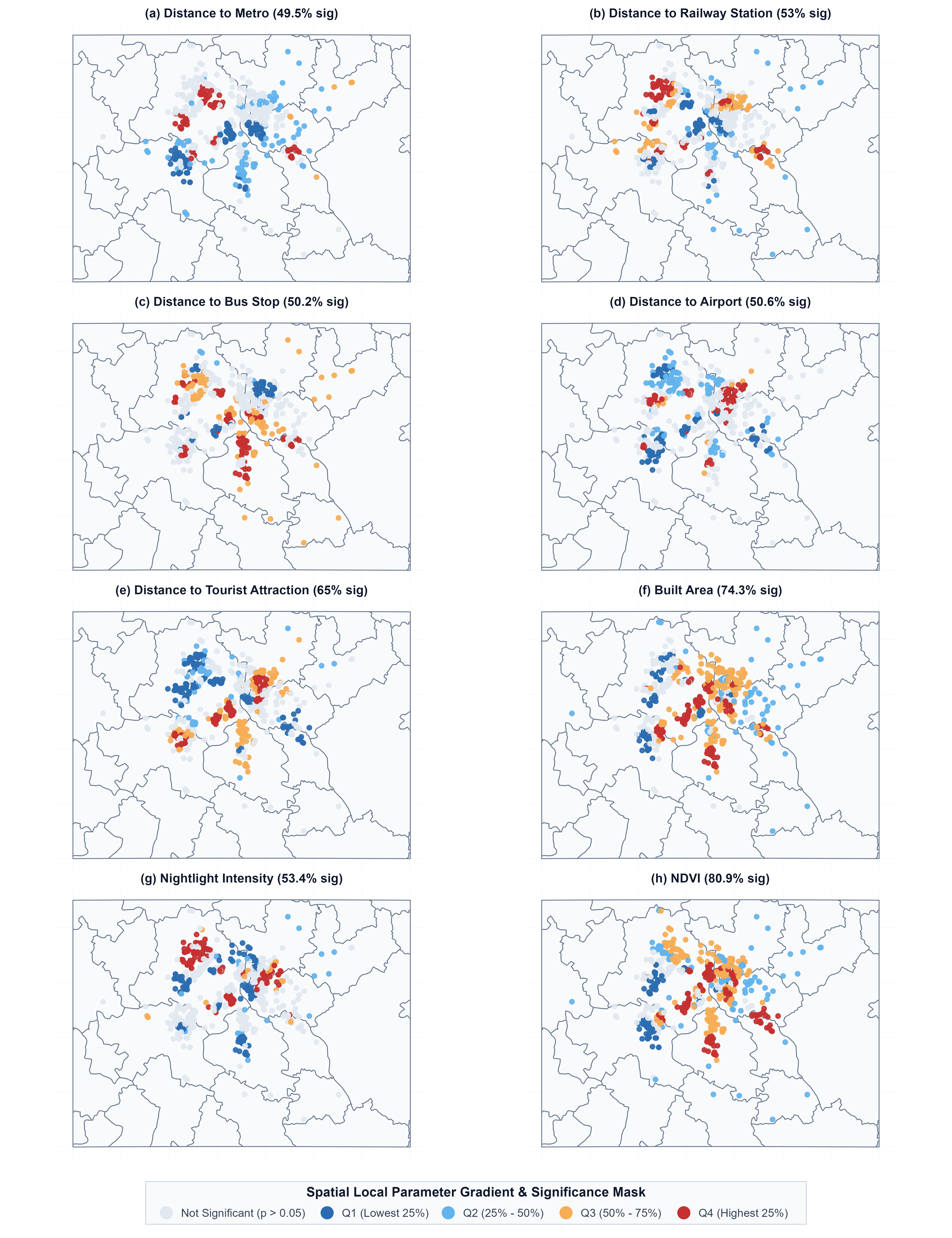}
      \caption{MGWR--SAR coefficient and significance maps for structural variables.}
    \label{fig:mgwrsar_structural_b}
\end{figure}
\restoregeometry
\clearpage

\subparagraph{Spatial Heterogeneity in the Effect of Night-time Light Intensity :} {\color{black}  

The coefficient surface, shown in Figure~\ref{fig:mgwrsar_environmental}(c)--(d) reveals substantial spatial variation in the effect of Night-time Light Intensity. Negative effects dominate in eastern and south-eastern clusters, indicating that higher urban intensity is associated with lower hotel prices in these areas. Conversely, positive effects in northern and certain central clusters suggest that greater economic activity and urban vibrancy enhances hotel prices. The significance map indicates that a substantial proportion of local coefficients are statistically significant ($38.04\%$), particularly within central and northern clusters. However, several peripheral areas display non-significant estimates, implying weaker or more uncertain effects of urban intensity in those regions. Overall, the results indicate that the valuation of urban intensity is context-dependent and geographically uneven.  \\

\subparagraph{Spatial Heterogeneity in the Effect of NDVI :} {\color{black} 

The NDVI coefficient surface in Figure~\ref{fig:mgwrsar_environmental}(e)--(f) also exhibits pronounced spatial heterogeneity, with both negative and positive effects observed across the study area. Clusters of negative NDVI coefficients are predominant in western and south-western locations, suggesting that higher vegetation density in these areas is associated with lower hotel prices. In contrast, positive and comparatively stronger coefficients are concentrated in central and eastern clusters, indicating that proximity to greener environments increases hotel prices in these sub-markets. NDVI is statistically significant in $86.23\%$ of regions, the second highest after discount percentage, highlighting its strong and widespread influence. Significance is particularly concentrated in central and northern zones, whereas peripheral areas display fewer significant estimates, reflecting variability in the strength and certainty of the vegetation effects. }

\section{Discussion and Conclusion}

This study employs the MGWR--SAR framework to examine the spatially varying determinants of budget hotel prices while accounting for spatial dependence and local environmental features. The results demonstrate that both localized spatial spillovers and environmental valuation (the green premium) play critical roles in price formation, and that relationships commonly assumed to be uniform in global hedonic models exhibit substantial geographic variation.

The local spatial autoregressive parameter $\rho(u_i,v_i)$ is significant in approximately half of the study area, confirming that price interactions between neighboring units exist but vary systematically across sub-markets. This spatial variation offers strong empirical support for spatial pricing dynamics, including coopetition and agglomeration theories. In core destination zones, positive spillovers reflect positive agglomeration externalities, where co-located hotels benefit from shared tourist footfall and demand. Conversely, in western and peripheral clusters, negative spillovers indicate price cannibalization or competitive undercutting, where hotels engage in intense price competition to capture highly price-sensitive budget travelers. Imposing a single global spatial parameter would mask these localized spatial dynamics.

Structural and reputational attributes also display clear spatial variation. While prior studies consistently report a negative relationship between Discount Percentage and price, and a positive association for Average Rating and Number of Amenities \citep{sainaghi2022effects, soler2019algarve, latinopoulos2018using, wang2019tourism, kim2020hotel, yang2025study}, our local findings show that the strength and even direction of these relationships vary across sub-markets. This suggests that global estimates conceal meaningful regional differences in how budget tourists value hotel reputation, scale, and promotions.

Locational factors similarly reveal transit-oriented pricing gradients that reflect a multi-modal tourist mobility portfolio. Proximity to the metro network acts as a primary proximity premium, reducing the spatial friction of navigating a dense mega-city. For budget tourists, the metro offers a cheap, rapid way to bypass urban traffic gridlock for daily sightseeing. In contrast, while proximity to railway stations is not globally significant, the local model reveals it carries a significant proximity premium in over 50\% of the study area, representing key inter-city arrival points for domestic travelers in those locations. Conversely, distance to bus stops exhibits a positive coefficient, representing a proximity discount. While bus stops provide transit access, they are also localized sources of noise, dust, and chaotic traffic congestion in Delhi NCR; consequently, tourists pay a premium to be located slightly further away in quieter streets. Proximity to the airport remains statistically insignificant, as the budget segment is less reliant on air travel gateways. While global models suggest uniform or non-significant effects, the local MGWR--SAR results prove that these accessibility spillovers are context-dependent and subject to localized transit dynamics. In peripheral or polycentric sub-cores (e.g., Gurugram), proximity to local transit hub infrastructure carries varying valuations depending on the surrounding neighborhood character and traffic congestion.

Environmental attributes, particularly vegetation density (NDVI) and Built-up Area, reveal a geographically differentiated green premium. The positive association of NDVI with price in central clusters demonstrates a green premium: in highly congested and dense urban cores, surrounding greenery is a scarce luxury that budget travelers value for its restorative, noise-reducing, and cooling benefits. Conversely, in peripheral areas, higher vegetation density is associated with lower hotel prices. In these outer zones, dense vegetation represents undeveloped land, indicating a lack of transport connectivity and safety concerns, which tourists discount. NDVI emerges as one of the most dominant and consistently significant variables, highlighting the key role of environmental quality in hotel price formation.

From a managerial standpoint, these findings suggest that pricing and marketing strategies should be location-specific rather than standardized. For budget hotel chains like OYO, which rely heavily on automated algorithms, integrating GIS-derived local spatial parameters (e.g., local transit proximity, green density) could help address "algorithmic pricing mismatches" and optimize localized yields. The strong green premium in urban cores also confirms the commercial value of urban greening, suggesting that operators can command price premiums by investing in local landscape aesthetics.

From a policy and planning perspective, the results emphasize the role of transport mobilities and environmental design in tourism destination development. Public transit networks do not generate uniform land-value premiums but interact with localized tourism clusters. Destination management organizations (DMOs) can use spatial coefficient mapping to plan green infrastructure, manage tourism corridors, and mitigate spatial overtourism by distributing budget travelers across connected sub-markets.

This study has several limitations. First, the analysis is limited to a single urban cluster (Delhi NCR), which may restrict generalizability. Second, the use of cross-sectional data prevents the examination of seasonal and temporal dynamics; future research utilizing panel data could capture evolving pricing patterns. Third, because the dataset consists of OYO Rooms listings, the prices reflect the centralized algorithmic rules of a single firm. While this offers an interesting perspective on algorithmic pricing behavior, future research should integrate independent budget hotels for comparison. Finally, the absence of occupancy and revenue data limits our ability to assess demand-side effects.

In conclusion, this study simultaneously addresses spatial dependence and spatial heterogeneity within a hedonic pricing framework for the budget hotel industry. By integrating advanced GIS variables, specifically vegetation density (NDVI), night-time lights, and built-up area, the study demonstrates the impact of environmental quality, transport mobility, and localized spatial interactions on hotel valuation. These insights are valuable for hospitality managers refining location-specific pricing and destination planners building sustainable tourism infrastructure.

\subparagraph{Declaration of interests:} {\color{black}
The authors declare that they have no known competing financial interests or personal relationships that could have appeared to influence the work reported in this paper. }

\subparagraph{Funding sources:} {\color{black}
This research did not receive any specific grant from funding agencies in the public, commercial, or not-for-profit sectors. }

\bibliographystyle{elsarticle-harv}
\bibliography{references}

\end{document}